\documentclass[acmsmall,nonacm]{acmart}

\renewcommand\footnotetextcopyrightpermission[1]{}

\usepackage{graphicx}
\usepackage{subcaption}
\usepackage{multirow}
\usepackage{booktabs}

\usepackage{enumitem}
\usepackage{listings}
\usepackage{xcolor}
\usepackage{verbatim}
\usepackage{url}

\definecolor{codegreen}{rgb}{0,0.6,0}
\definecolor{codegray}{rgb}{0.5,0.5,0.5}
\definecolor{codepurple}{rgb}{0.58,0,0.82}
\definecolor{backcolour}{rgb}{0.95,0.95,0.92}

\lstdefinestyle{mystyle}{
    backgroundcolor=\color{backcolour},   
    commentstyle=\color{codegreen},
    keywordstyle=\color{magenta},
    numberstyle=\tiny\color{codegray},
    stringstyle=\color{codepurple},
    basicstyle=\ttfamily\footnotesize,
    breakatwhitespace=false,         
    breaklines=true,                 
    captionpos=b,                    
    keepspaces=true,                 
    numbers=left,                    
    numbersep=5pt,                  
    showspaces=false,                
    showstringspaces=false,
    showtabs=false,                  
    tabsize=2
}

\usepackage{soul}
\usepackage{todonotes}
\presetkeys{todonotes}{inline}{}

\AtBeginDocument{%
  \providecommand\BibTeX{{%
    \normalfont B\kern-0.5em{\scshape i\kern-0.25em b}\kern-0.8em\TeX}}}

\begin{document}

\title{Living Databases: A Unified Model for Continuous Schema Evolution, Versioning, and Transformations}

\thanks{%
© 2026 IEEE. This is the author's version of the work accepted
for presentation at the 42nd IEEE International Conference on
Data Engineering (ICDE 2026), Data Engineering Future Technologies (DEFT) track. The final version of record will be
published in the ICDE 2026 proceedings and will be available via
IEEE Xplore.%
}

\author{Amol Deshpande}
\affiliation{%
  \institution{University of Maryland}
  \city{College Park}
  \state{Maryland}
  \country{USA}
}
\email{amol@umd.edu}

\begin{abstract}
Databases, and datasets more generally, evolve continuously through updates, transformations, versioning, schema changes, streaming operations, and other mechanisms. While prior work has noted connections among some of these areas, they have traditionally been studied in isolation, each with its own abstractions, algorithms, and system implementations. In this paper, we argue for unifying these diverse functionalities under a single abstraction and a common set of computational primitives. We present such an abstraction, powerful enough to encompass existing use cases and to support new ones. Going beyond previous approaches, our framework seamlessly integrates provenance tracking for system-visible operations, conditional propagation of updates, and configurable alerts on change events.  It also offers a principled treatment of dependent objects such as views and derived artifacts like machine learning models, by providing declarative mechanisms to control their evolution. Finally, we sketch a prototype implementation in a relational-like database system based on an adaptation of the \emph{Prolly Tree}, a Merkle tree–inspired data structure with tunable parameters to meet varying performance requirements, and present some initial experimental results.

\end{abstract}

\maketitle


\section{Introduction}
The modern data landscape is defined by constant evolution.  Datasets undergo continuous transformation, update, and reinterpretation to support diverse applications and analyses.  In database systems, several fundamental challenges arise from this dynamic nature, including:

\vspace{4pt}
\noindent\textbf{(1) \underline{Data Versioning:}}  
Although version control systems like {\em git} have become standard in software development—and despite clear need for similar abstractions in data
management~\cite{datahub2015,bhattacherjee2015principles}—robust support for data versioning in database systems remains limited.  While support for temporal version chains (e.g., Oracle Flashback Time Travel, Delta Lake~\cite{armbrust2020delta}) is common, many use cases demand branching, merging, and isolated environments\footnote{\url{https://neon.com/branching}}.  These include:  
\begin{enumerate}[label=(\alph*)]
\item sandboxing for development, testing, or schema migrations without impacting production;  
\item maintaining alternate “what-if” realities for scenario planning;  
\item collaborative editing and sharing;  
\item generating PII-sanitized or anonymized copies for compliance (e.g., GDPR, HIPAA);  
\item machine-learning experimentation;  
\item offline or distributed workflows~\cite{gray1996dangers,yilmaz2025generic}; and  
\item agentic workflows~\cite{adityapaper}.  
\end{enumerate}
In addition to supporting the creation and maintenance of a large number of branches efficiently, other key requirements include configurable merge semantics, and efficient cross-branch operations (e.g., \emph{diff}, \emph{blame}). From the performance perspective, copy-on-write cloning and cross-version similarity detection and compression are essential to reduce the storage overheads and the latencies of creating new branches.

\vspace{4pt}
\noindent\textbf{(2) \underline{Schema Evolution:}}  
Schema evolution is both a necessity for databases to handle evolving user requirements, as well as a constant pain point for decades~\cite{stonebraker2016database}. 
Despite significant research on this topic (which we cover in more detail later), almost no commercial database
supports schema evolution in a flexible, general, and transparent manner. In most cases, schema evolution and the associated issue of data migration are 
handled by layers on top of the database (e.g., object-relational mappers like Django), which are known to suffer from significant performance issues. 
Some key requirements here include instantaneous schema changes and the ability to run applications against multiple versions of schemas. 

\vspace{4pt}
\noindent\textbf{(3) \underline{Materialized Views:}}  
Materialized views have been a staple in databases for a long time, and many incremental view maintenance techniques have
been developed over the years to efficiently propagate base table updates to the materialized views~\cite{budiu2023dbsp}. However, a few new requirements have emerged in recent 
years, including scheduled updates as well as conditional or no updates (e.g., based on {\em
freshness}\footnote{\url{https://docs.snowflake.com/en/user-guide/dynamic-tables-about}}). These are often built on top of the databases, through use of triggers or cron jobs. 
There are several other related concepts, including {\em continuous queries over streaming data}, {\em data transformation (e.g., ETL) pipelines}, {\em data caching}, etc. 
To the extent that these operations live inside the same data management system, the issues therein are similar to materialized views. 

\begin{figure}[t]
  \centering
  \includegraphics[width=0.85\linewidth]{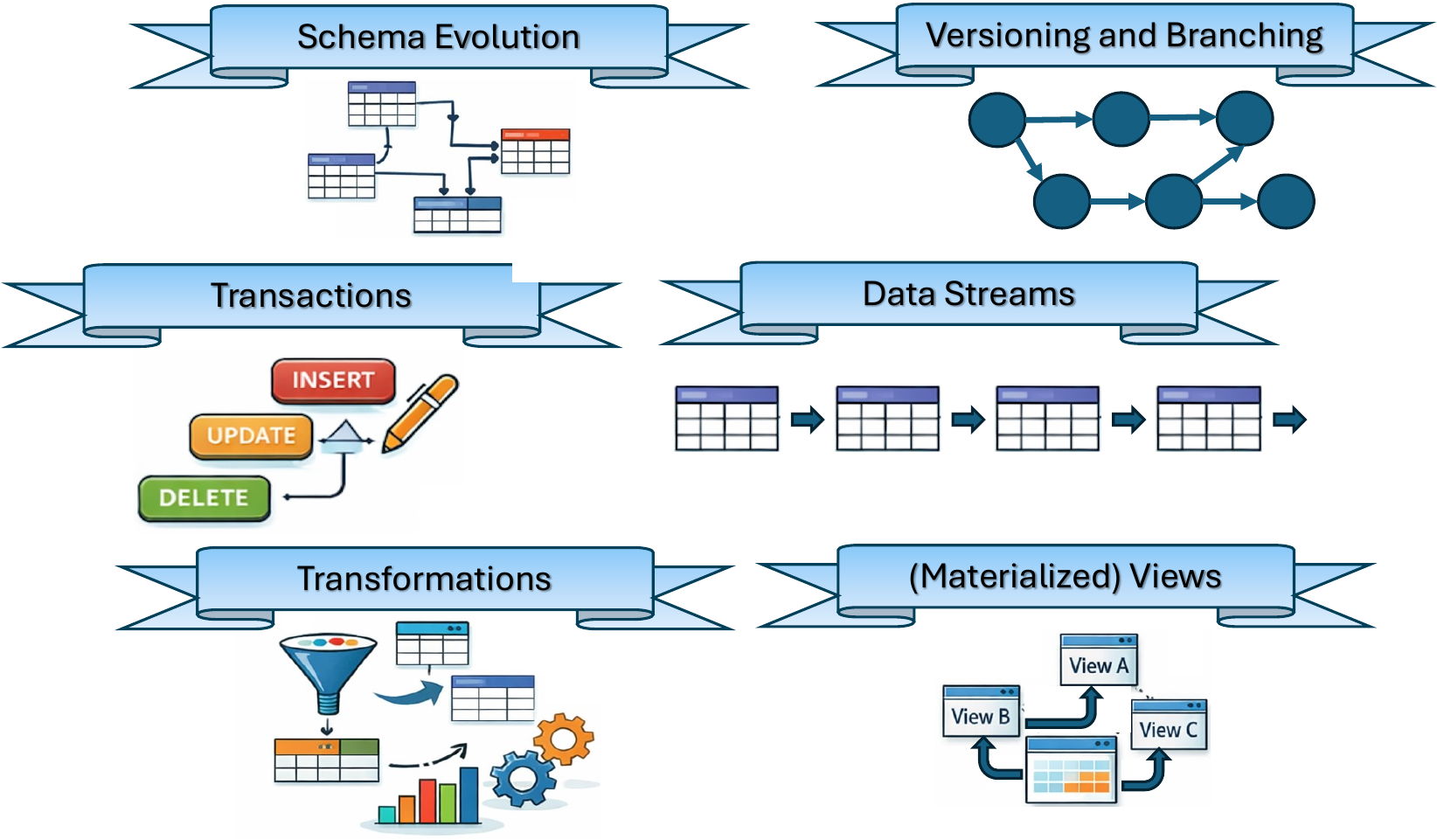}
  \caption{Our proposed abstraction aims to unify several traditionally distinct functionalities}

 \label{fig:high-level}
\end{figure}

The key motivation for our work is \ul{the observation that, historically, these topics have been treated as distinct research areas, each with its own specialized abstractions, algorithms, and system implementations}.  A closer examination, however, reveals many similar ideas and a shared set of computational primitives.  At their core, all of these topics concern the derivation of one dataset from another: data transformation systems explicitly map input datasets to output datasets; version control systems manage a history where each version is a transformation of its predecessor; schema evolution transforms the entire dataset to fit a new schema; and materialized view maintenance keeps a derived dataset consistent with its base tables through incremental updates.  Common techniques—incremental computation, lazy materialization, and efficient differential analysis—recur across these domains, suggesting that their separation is more historical than fundamental.

\medskip
In this paper, we argue that these seemingly disparate concepts should be unified under a single, cohesive framework.  We introduce \ul{a generalized abstraction for data derivation that captures the core requirements of transformation, versioning, evolution, and view maintenance, while also enabling novel, synergistic capabilities not attainable in individual systems}.  

Our proposed abstraction is built around the notion of a {\bf snapshot graph}, which represents database snapshots over time as a directed acyclic graph—much like a {\bf version}
graph~\cite{dex} or a {\bf commit} graph\footnote{We chose this term instead of those existing terms, or alternatives like {\bf derivation} graph, because of
the existing connotations of those terms.}.  Snapshots form linear {\bf branches} that may overlap; updates are allowed only at branch {\em heads}, though new branches may originate from any existing snapshot. Furthermore, each
edge carries rich metadata describing the derivation operation, provenance, and propagation conditions (e.g., unidirectional vs.\ bidirectional, conditional triggers, delays), as we discuss in more detail in the next section.

A natural question is: why unify these concepts?  Beyond eliminating duplicated ideas, our framework enables cross-cutting solutions that extend far beyond the basic functionality currently supported.  We highlight a few benefits here and discuss others throughout the paper:

\begin{enumerate}
  \item {\bf Cleaner treatment of dependent objects:}  When a base table undergoes a schema change or other alteration that would invalidate a view (or another derived object), our model lets one defer view updates so dependent applications can continue running without interruption.
  \item {\bf Versioning of dependent objects:}  In existing systems, dependent artifacts (e.g., views) are not versioned alongside their base tables—leading to potential
inconsistencies.  We naturally version all derived objects, and provide the ability to retrieve a specific past version of such an object.
  \item {\bf Provenance:}  By elevating lineage to be a first-class citizen, the system not only performs transformations but can also explain them, which is critical for debugging,
auditing, compliance (e.g., GDPR), and understanding complex pipelines~\cite{miao2017provdb}.
  \item {\bf Conditional propagation:}  Rich edge metadata makes it straightforward to specify conditional or delayed updates -— functionality that is cumbersome or impossible under most current abstractions.
\end{enumerate}

To ground our vision in practice, we describe the design and prototype implementation of this abstraction within an entity–relationship database system from our prior work~\cite{erbium}.  While the abstraction is broadly applicable, we rely on concrete definitions of schemas and transactional updates to ground our discussion 
and to guide our prototype implementation.

\section{Abstraction}
The proposed abstraction is centered around the notion of a {\bf snapshot} graph, which is a directed acyclic graph with {\bf snapshots} as nodes. A snapshot (called a
{\bf version} or a {\bf commit} in similar contexts) is a point-in-time instance of a database, (in our case) with an associated {\bf schema}.  Unlike typical version or commit graphs, our abstraction is significantly richer and attempts to capture additional context, operations, and propagation semantics. To simplify the exposition, we first briefly recap the concepts of versioning and branching, and then discuss the additional annotations supported by our abstraction.

\begin{figure}[t]
  \centering
  \includegraphics[width=.75\linewidth]{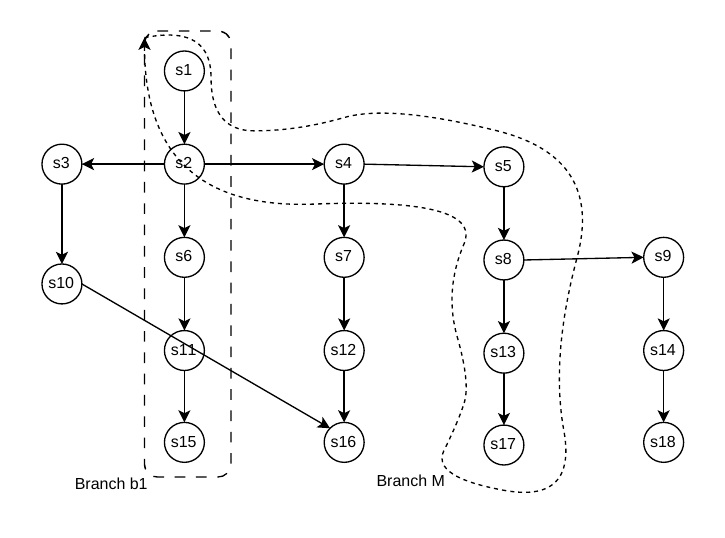}
  \caption{Example of a snapshot graph without annotations (only two branches, $M$ and $b_1$, are explicitly shown)}
 \label{fig:version_graph_only}
\end{figure}

\subsection{Snapshots and Branches}
Figure \ref{fig:version_graph_only} shows an example snapshot graph with 18 snapshots, each with an associated schema and table instances. A {\bf branch} corresponds to a linear chain of snapshots that captures the evolution of a database through either a sequence of {\bf transactions} or {\bf schema modification operations}.
Branches have externally visible names that are used by users and applications to {\em attach} to a branch, i.e., to read the contents of the branch or to modify it. The {\bf head} of a branch refers to
the latest snapshot in the branch, and is the only one against which DML transactions can be initiated.

Two branches are explicitly shown in the example. {\bf Branch M} consists of snapshots $s_1 \rightarrow s_2 \rightarrow s_4 \rightarrow s_5 \rightarrow s_8 \rightarrow s_{13}
\rightarrow s_{17}$, with $s_{17}$ as the head; whereas {\bf Branch b1} consists of five snapshots, $s_1, s_2, s_6, s_{11}, s_{15}$. The two branches share the first two snapshots, and diverge after that, with other branches being created along the way.

The figure also shows an example of a {\bf merge} where snapshots $s_{10}$ and $s_{12}$, from two different branches, are merged into a single snapshot $s_{16}$. The specifics of how the merge operation works is orthogonal to the snapshot graph abstraction. Merge may be done as a symmetric operation ($s_{16}$ belongs to both branches) or as an asymmetric operation ($s_{16}$ only belongs to one of the two branches, and the other branch continues on from the previous snapshot).

\subsection{Capturing Context through Annotations}
\begin{figure}[t]
  \centering
  \includegraphics[width=.75\linewidth]{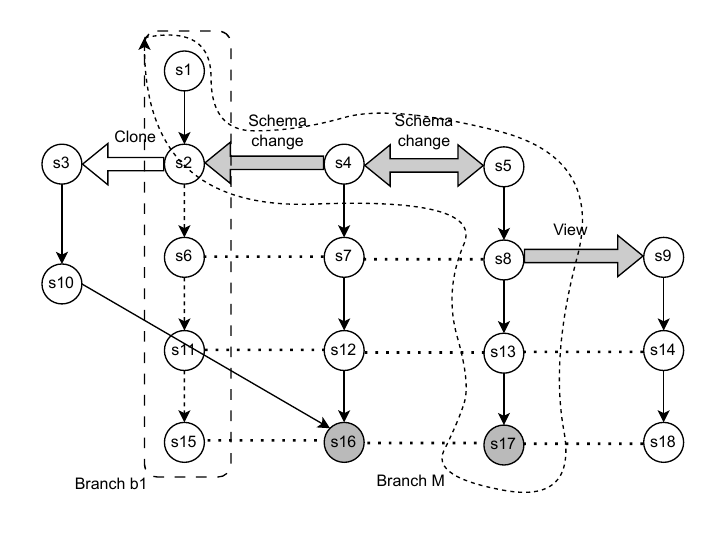}
  \caption{Example snapshot graph with annotations (only two branches, $M$ and $b_1$, are explicitly shown)}
 \label{fig:snapshot_graph}
\end{figure}

The abstraction discussed so far is similar to how most versioning systems like {\em git} work. Next, we discuss how we enrich this abstraction with different types of edges and annotations (Figure \ref{fig:snapshot_graph}), to unify the treatment of transactions, schema evolution, data streams, and dependent objects like views.

The enriched snapshot graph contains a few different types of edges. 
\begin{itemize}
  \item The edges denoted by solid arrows capture DML transactions, i.e., inserts/updates/deletes to the database as well as any ``merge'' operations (as noted above, these can only be done against the branch heads, shown as shaded nodes in the figure). 
  \item A metadata change, or creation of a view or other dependent object, or a {\bf clone} operation, is captured through creation of a new branch, and the corresponding edges are shown as block arrows in Figure \ref{fig:snapshot_graph}. For some of these changes, e.g., a schema evolution, the branch name may follow the change (e.g., {\em branch M} in Figure \ref{fig:snapshot_graph}).
\item The dashed lines (e.g., $s_6 - s_7$) show automatic updates as discussed below.
\end{itemize}

We note that most of these are for internal bookkeeping purposes, and do not need to be visible to the users or applications (who primarily work with branches).
The block arrows have additional annotations that drive key functionalities in our abstraction.

\begin{figure}[t]
  \centering
  \includegraphics[width=.75\linewidth]{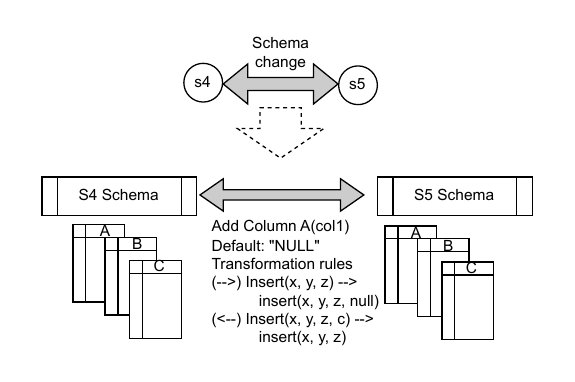}
  \caption{Detailed Example of a Schema Change}
 \label{fig:detailed_example}
\end{figure}

\vspace{4pt}
\noindent\underline{\bf Sync Semantics:}
First, the block arrow may indicate whether the two endpoints should be ``kept in sync'', and whether that sync is bidirectional or uni-directional. The shaded block arrows in
the figure indicate a sync, and the directionality of the arrow indicates the behavior. For example, the second schema change (shown in more detail in Figure
\ref{fig:detailed_example}) is annotated with a bidirectional sync, i.e., either of the two branches can be updated and those updates will be propagated to the other
branch. This has been observed as a key requirement for transparent schema evolution where all the applications need not be immediately updated to use the new schema, 
and instead can be lazily moved over, reducing the downtime~\cite{curino2013automating,rae2013online,bhattacherjee2021bullfrog,herrmann2018multi,Hu2023Tesseract}. 
Such a bidirectional sync can also capture an {\em updateable view}. 

Of course, bidirectional sync is only allowable for a limited set of schema changes, as noted in that prior work. For example, addition of a new column with a default value is
compatible with bidirectional sync, as well as a deletion of a {\em nullable} column. Similarly, a schema change where two tables connected by a foreign key are merged into a
single table can be supported as long as the foreign key is nullable. We refer to the prior work for a more elaborate discussion of compatible schema changes.

For some schema changes, only reverse sync may be possible ($s_4 \rightarrow s_2$), e.g., addition of a column that does not have a default value and is not nullable. We can, in effect,
define a view on the new schema that is identical to the old schema; then any read-only applications (e.g., dashboards) can continue to work until they are migrated. Finally, for
schema changes where no sync is possible, our abstraction allows for the possibility that a read application continues to run against the last compatible snapshot of the data.

We also support operations like {\em clone} (also called {\em fork}) without any requirement to keep the two branches in sync (shown as an unshaded block arrow). As 
discussed in the previous section, this is 
a very common operation in collaborative data analytics~\cite{datahub2015}, and forms a major motivation for supporting versioning as a first-class construct in databases. 
It is also common to see a {\em merge} from such a cloned branch back into the original branch, and those can be seen as DML updates against the target branch. If the {\em merge}
requires a schema change in the target branch, it would be captured as a sequence of two operations, i.e., a schema change followed by the merge.

For dependent or derived objects like views (e.g., $s_8 \rightarrow s_9$), unidirectional sync is always defined as long as there is no incompatible metadata change. 

\vspace{4pt}
\noindent\underline{\bf Conditional Propagation:}
However, this brings us to another key construct in our abstraction, i.e., {\bf conditional propagation}. A sync edge may be annotated with one or more conditions that dictate
which changes to the source should be propagated to the target. Although this makes most sense for derived objects like views, it can also be useful for schema changes. 
Some examples of such conditions include a metadata change involving specific tables (that may be incompatible with the target branch), or the relative size of the change (indicating a major change to data characteristics that needs to be reviewed).

If the condition is not satisfied for a change, then the sync must be {\bf disassociated} and no further changes should be propagated.
In some cases, such a condition may be coupled with an {\bf alert} that is raised to an administrator to take further action. Development of a comprehensive set of conditions,
and efficient implementation of them, is a key open problem here.

\vspace{4pt}
\noindent\underline{\bf Propagation Frequency:}
Finally, unidirectional sync/propagation edges can also be annotated with {\bf when} the changes should be propagated. Specifically, we propose supporting the standard 
constructs from triggers, namely, {\bf immediate} vs {\bf deferred} (i.e., whether the propagation should be within the same transaction or not), {\bf on-demand}, or {\bf periodically} (e.g., only propagate every hour). Different variations of these are supported for views and triggers in different database systems, with periodicity often handled 
through external cron jobs; this abstraction allows us to unify these concepts as well as their implementation.

We note that these constructs (including propagation logic) are {\em all} logical constructs, and do not dictate when physical data structures are modified. For example, 
a view may or may not
be materialized, which would have an impact on when the computations are done. Similarly, a data migration due to a schema change may be done lazily in the background, even if
logically the two branches need to be kept in sync. These decisions of course have a significant impact on the performance, as we will discuss in the next section.

\subsection{Correctness}
The above discussion and requirements can be synthesized into a set of criteria that dictate whether a snapshot graph is {\em correct} and should be allowed, which in turn
impacts which operations are allowable.
\begin{enumerate}
\item Only the {\bf head} snapshots in a branch can be updated. As discussed in recent work, the ability to {\bf merge} branches can be used to support concurrent transactions
in a transparent manner~\cite{yilmaz2025generic}.
\item A branch that is the target of a unidirectional sync (e.g., {\em Branch b1} in Figure \ref{fig:snapshot_graph}) can not be updated without disassociating it from the sync (i.e., without removing that sync).
\item A branch that is the target of a unidirectional sync cannot be involved in a bidirectional sync.
\end{enumerate}
The last condition above ensures that we can always propagate updates across branches in a consistent manner, although we expect chains of synced edges to be quite uncommon due
to the performance implications.

\subsection{Is the Abstraction Too Complex?}
A natural concern here is that the snapshot graph abstraction above may be too complex for users to use effectively. We argue that 
the abstraction is not substantially more difficult to understand and use than version control systems like {\em git} with which most users are quite
comfortable today. Although there will be some initial learning curve, we believe that the abstraction will prove cleaner to use over the long term, especially compared to 
the hodge-podge of constructs
that the users need to work with today. It will especially reduce the use of out-of-the-band mechanisms for dealing with issues such as maintenance of derived objects
or data migrations, by bringing those inside the database and providing visibility into them. 

More specifically, we expect most users or applications will be attached to a single branch, and will issue their read/write/update queries against that branch.
For breaking schema changes, the decisions about how the evolve the applications will need to be made by the application maintainer. In some cases, if the original and new
schemas can co-exist, then the abstractions above can help the users or the application maintainer navigate such changes. Similarly, if a view dependent on a branch needs
to be disassociated from that branch (due to a breaking metadata change), a read application (e.g., a dashboard) can continue to run while a new view is being 
defined and the application is migrated appropriately. Since all of these operations are recorded in the database, it is easier to keep track of them, take 
appropriate actions, and audit them. 

Furthermore, the ability to set alerts on the syncs can help flag any issues in a systematic and uniform manner. This can be particularly useful in collaborative settings where
many users are concurrently making changes to the database, and they don't all have visibility into the actions being taken by others. By continuously monitoring the propagation 
of changes across derivations or transformations or views, a number of issues can be identified in time and appropriate actions can be taken.

\section{Implementation Considerations and Prior Work}
The abstractions we propose above are quite rich and complex, and performance is naturally a major question mark. 
Here, we briefly discuss some of the key considerations and tradeoffs through a review of closely related prior work in a number of different areas.
In the next section, we present our ongoing prototype implementation and discuss how it addresses those considerations.

To make our discussion here concrete, we assume an operational database that is being transactionally updated in a continuous manner, with a reasonable mix 
of complex analytical queries that need to be executed efficiently (i.e., the setting where systems like PostgreSQL shine). Some of the key motivations 
underlying our work, including the need to support schema evolution without downtime, incremental maintenance of derived objects like views, and user-driven versioning, are
amplified in this setting. Although most of the discussion here is more general and applies to other settings such as data
warehouses as well, some of the specific data structures we discuss wouldn't make sense in that setting. 

At a high level, the main performance considerations are: 
\begin{enumerate}
    \item Storage space required to store the large number of snapshots that may be created implicitly or explicitly.
    \item Transaction throughput in the presence of synced branches.
    \item Performance for read-only analytical queries that may scan large volumes of data.
    \item Efficient propagation of updates.
    \item Latency of major operations like schema evolution or cloning.
\end{enumerate}

\vspace{2pt} \noindent\underline{\bf Versioning and Deduplication:}
There has been an extensive body of work on {\bf storage deduplication}~\cite{paulo2014} and {\em dataset versioning} that has looked at some of these issues, especially the
storage space vs retrieval latency tradeoff~\cite{bhattacherjee2015principles,dex}. The primary objective of the work on storage deduplication is achieving high ingest rate as well as high
compression (given they are often dealing with data rates in excess of 100MB/s), and retrieval is a secondary concern. Most of the techniques work by partitioning the data into
{\bf chunks} of varying sizes, and identifying and eliminating duplicate chunks. The approach we discuss in the next section uses hash-based boundary detection and variable-sized
chunks, which is also commonly seen in the work here. Small updates to the data (i.e., transactions) are not a commonly supported operation. 

Source code versioning systems like {\em git} as well as several recent works on data versioning adopt the abstraction of a {\em version
    graph}~\cite{datahub2015,huang2017orpheusdb,decibel,bhattacherjee2015principles,wang2018forkbase}, and propose different data
structures or techniques to exploit the commonalities across versions. {\bf Delta encoding} is commonly used for source code as well as unstructured datasets, wherein some versions (snapshots) are materialized in their entirety, and others are maintained as {\em deltas} from those. Although there have been some attempts at supporting granular querying over such delta-encoded versions~\cite{dex}, generally speaking, this approach is most suitable for retrieving or querying entire versions. Decibel~\cite{decibel} 
investigated versioning in a relational database, and proposed a hybrid solution that combines delta encoding with a {\bf bitmap-based} approach that tracks which tuples exist in which versions. This is similar to the approach taken in OrpheusDB~\cite{huang2017orpheusdb}, which ``bolted on'' versioning support in PostgreSQL. For bitmap-based approaches, the metadata (i.e., tracking which tuples belong to which versions) can get very large and can be difficult to maintain and manipulate efficiently. 
Inspired by blockchains and Merkle trees, Forkbase~\cite{wang2018forkbase,yue2020analysis} uses {\em content-based slicing} to create a B+-tree-like index that naturally exposes sharing 
opportunities between versions. We discuss a variation of this, as implemented by the open-source systems Noms and Dolt, in the next section. 
All of these support efficient transactions compared to the techniques discussed above, although they are more expensive than using vanilla B+-trees due to the additional hash computations required.

\vspace{2pt} \noindent\underline{\bf Schema Evolution:}
Schema evolution is an unavoidable consequence of the application development lifecycle, due to changing user requirements, bug fixes, 
and performance optimizations, and have been supported in database DDLs since the beginning. However, even today, applying schema changes 
tends to be a disruptive and offline process, requiring significant system 
downtime and manual, error-prone data migration scripts. The applications running on top of the database often need to be migrated as well~\cite{vassiliadis2023joint}. 
This pushes the developers towards making small, incremental changes to the schema
that often result in schemas that are unintuitive and not normalized (called {\em database decay}~\cite{stonebraker2016database}). 
There has been increasing interest~\cite{curino2013automating,rae2013online,bhattacherjee2021bullfrog,herrmann2018multi,Hu2023Tesseract} in recent years in 
supporting schema evolution without application downtime, by: (a) letting multiple 
versions of the schema co-exist and (b) lazily and transparently migrating the data from the original schema to the new schema. 
F1~\cite{rae2013online} is a distributed key-value store that allows applications running against two different schemas to co-exist, by treating the two schemas as views
against the same database (this is possible because of their relatively simple data model, which doesn't really require a ``migration'').
BullFrog~\cite{bhattacherjee2021bullfrog} supports instantaneous schema changes (from user perspective) by migrating the data from the old schema to the new schema on-demand 
(in response to user queries or transactions); it also allows applications to run against the old schema for backward-compatible schema changes. 
InVerDa~\cite{herrmann2018multi} is a multi-version database where schema changes are written in a bidirectional database evolution language (BiDEL), conceptually similar
to bidirectional syncs in our abstraction; they don't consider schema changes that cannot be written using BiDEL. 

InVerDa also considers physically materializing a subset of the versions for efficiency. Although this leads to better performance of read queries, it makes 
it harder to support serializable transactions since a transaction against one schema, in effect, requires a two-phase commit across two or more materializations. This forms 
one of the key performance-related challenges in such a system (including ours). Finally, Tesseract~\cite{Hu2023Tesseract} treats a schema change as a change to the entire database, and uses a relaxed {\em snapshot isolation} protocol to allow concurrent DDL and DML transactions. 

\vspace{2pt} \noindent\underline{\bf View Maintenance:}
There is also a large body of work on incremental view maintenance, where many algorithms and auxiliary data structures have been developed to efficiently propagate
base table changes (deltas) to the views (a recent work, DBSP~\cite{budiu2023dbsp}, can be seen as unifying and generalizing many of those techniques in a single framework). 
Those techniques can be adapted as is in our abstraction if full materialization is desired; however, partial materialization, which can be a powerful feature and is 
naturally supported in our abstraction, has seen limited work~\cite{luo2006partial}.

\vspace{2pt}
Some other performance considerations have been less studied. For example, {\em garbage collection} becomes a critical question as the number of historical snapshots 
increases over time. There are both implementation as well as conceptual questions here. Semantically, the snapshot graph abstraction requires the ability to retrieve
any past snapshot, which limits the opportunities for garbage collection; so additional constructs may need to be developed to specify the behavior for older snapshots.
From implementation perspective, the system could compress the older versions more, thereby reducing the storage taken by them but resulting in increased latency.

Finally, to support complex analytical queries, we believe the storage layer should support vectorized storage and computations. We are not aware of prior work that has looked
at this in the context of data versioning or schema evolution.

\section{Prototype Implementation}

To realize the snapshot graph abstraction, we are developing a new storage engine designed for the efficient management of evolving, overlapping snapshots. Our implementation adopts a {\em Prolly-tree-based storage structure} with a {\em columnar layout}, grounded in the principles of {\em content-addressable storage}~\cite{wang2018forkbase}, but with several key architectural innovations to support the unique demands of our model. 

We chose this design as a starting point for several reasons. First, the probabilistic chunking used in Prolly trees (as discussed in more detail below) ensures that small updates affect only a limited portion of the structure, allowing unchanged regions to be reused across snapshots through structural sharing. This enables efficient management of large numbers of related snapshots while automatically maximizing storage reuse. Second, unlike delta-based approaches that represent snapshots as chains of differences, the Prolly-tree organization avoids long reconstruction paths and supports updates and transactional modifications more naturally. Third, the key-ordered tree organization naturally supports efficient lookup of individual tuples (by key).

Our design further introduces three key extensions to this basic structure. First, we support flexible chunking policies that allow the system to fall back to more deterministic, B+-tree–like chunking when appropriate, reducing overheads in scenarios where updates are sparse or well-localized. Second, we adopt a columnar layout within chunks to improve compression opportunities and enable efficient scans for analytical workloads. Third, the system supports lazy evaluation, where chunk contents may be defined (as operations on other chunks) without immediately materializing the corresponding physical data, allowing the system to defer work until it becomes necessary and to better amortize materialization costs across operations.

The combination of structural sharing, key-based navigation, and columnar storage provides a unified representation that supports both transactional updates and analytical access patterns efficiently. Finally, the system is designed so that most of the storage layout and materialization decisions—such as chunking policy, chunk sizes, compression strategies, and physical clustering—can be managed by an autonomous tuning engine that transparently optimizes for performance and storage efficiency based on observed workloads.

\subsection{Background: Prolly Trees}
Prolly Tree is short for ``Probabilistic B-tree'', a data structure originally developed for the Noms database\footnote{\url{https://github.com/attic-labs/noms}}, and also used
by the Dolt database\footnote{\url{https://github.com/dolthub/dolt}}. Prolly trees are structurally very similar to B+-Trees, with the main difference being that the nodes are split
based on their content rather than size.
Consider a B+-tree containing $n$ keys, $k_1 \le \cdots \le k_n$, and let $h$ denote a hash function that takes as input a sequence of keys. Further let $W$ and $C$ be
pre-determined constants. The boundary of the first leaf
node will be the smallest $i$ such that $h(k_{i-W+1}, \cdots, k_i) < C$. The two constants ($W$ and $C$) can be used to tune the structure of the resulting tree. 
For each leaf, another hash function is used to compute an {\em
address} for the leaf node based on the entire content of the node (i.e., keys as well as values); the content address is used to locate the leaf node in the storage (which can
be thought of as a {\em map}) and is also used as the key for the next level of the Prolly Tree. Similarly to B+-Trees, the same process is repeated for the second level to create
interior nodes, and so on until a single root node is created (again based on the probabilistic splitting).

\vspace{2pt}
\noindent \underline{\bf History Independence:} A main advantage of the Prolly tree over a B+-Tree is that the content-based chunking naturally exposes the sharing opportunities across the 
snapshots. This is called {\em history independence}\footnote{\url{https://docs.dolthub.com/architecture/storage-engine/prolly-tree}}, which is understood to
mean: the structure of the Prolly tree is independent of the order in which the keys are inserted into the index, i.e., it only depends on the actual content (keys and values).
This is true by definition for the original Noms proposal, which states that ``[t]o mutate a Prolly Tree, conceptually we build a new Prolly Tree from scratch...''.
However, the Dolt implementation performs incremental updates rather than conceptually rebuilding the tree from scratch. While the chunking rules are content-based, it is not immediately clear from the documentation whether the resulting structure is strictly canonical (i.e., identical to a full rebuild) under all insertion orders. In particular, localized splitting decisions may affect the boundaries of subsequent leaf nodes, potentially leading to structural differences depending on the mutation sequence. We plan to investigate this empirically in future work.

Despite this issue, the Dolt implementation is simpler from the concurrency perspective and forms the basis of our implementation. We are planning to further investigate the
properties of Prolly trees, and in particular, how much sharing they enable across similar snapshots, especially compared to the storage deduplication approaches, through 
a comprehensive experimental evaluation in our future work.

\subsection{Adaptable Columnar Content-based Indexes}
Our storage layer is built around a Prolly tree-like index that partitions relations into immutable, content-addressed chunks in a columnar fashion. 
Unlike traditional Prolly Trees which use a fixed rolling hash function to determine chunk boundaries, our approach allows for more \textbf{flexible chunking policies}. 
Specifically, 
the system also supports simpler, B+-tree-like policies where chunks 
are split when they reach a specific size. This flexibility is the first step towards an adaptive system, where the chunking strategy can be chosen on a per-relation 
basis based on the access patterns. 

Within each chunk, data is stored in a \textbf{columnar format}, with or without {\bf attribute splitting}. Attribute splitting refers to vertical partitioning of a single 
relation (each partition must contain the primary key) and building a separate index on each set of attributes (similarly to C-Store ``projections'').
For analytical queries that perform large table scans, columnar storage provides significant performance advantages through improved cache utilization and compression. 
This also benefits 
transformations and schema evolution operations that often only affect a subset of attributes. 

To capture lazy or on-demand materialization or migration, our design also supports the notion of {\bf virtual chunks}. 
This means that a new snapshot created via a branch or schema evolution
operation can be instantiated instantaneously at the metadata level, but the chunks for the new snapshot are not immediately materialized. Instead, they are created as 
``virtual chunks'' that store a recipe for their computation---pointers to the source chunks and the transformation logic to be applied. 
Data is only physically generated when a chunk is accessed for the first time (a policy that migrates in the background in periods of low load can also be supported). 
This mechanism makes branching and many forms 
of schema evolution zero-cost operations from a user's perspective. 

\begin{figure}[t]
  \centering
  \includegraphics[width=0.81\linewidth]{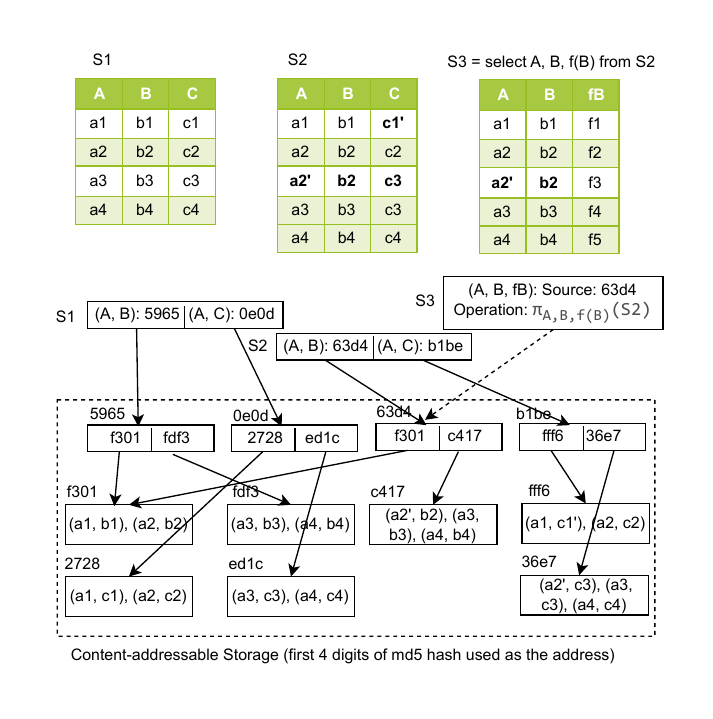}
  \caption{Content-addressable Storage and Indexes}
 \label{fig:cas}
\end{figure}

Figure~\ref{fig:cas} shows an example with three snapshots ($S1, S2, S3$) that share some tuples and columns. For example, $S1$ is vertically split into two attribute groups ($A$ is the primary key), each of which spans 2 leaf chunks (and one interior chunk). The same is true for $S2$, with one chunk shared with $S1$. On the other hand, $S3$ is only maintained as a virtual chunk, with a pointer to the source chunk ({\em 63d4}). When $S3$ is accessed, the system may decide to materialize it using the source chunk (which in turn would require reading the corresponding leaves), or it may temporarily compute the chunks but not materialize them in the CAS.

The flexibility inherent in our design---variable chunking policies, vertical partitioning, and configurable materialization strategies (lazy vs. eager)---creates a vast and complex optimization space. 
Therefore, a key component of our long-term vision is an \textbf{autonomous tuning engine}, that will monitor query and update workloads on the snapshot graph, to make transparent decisions about the physical data layouts. It will decide when to materialize a virtual chunk, which attributes to split into separate chunk trees, and what chunking policy to apply, all with the goal of dynamically adapting the system's physical configuration to best serve the observed access patterns.

\subsection{Preliminary Experimental Results}
We are building an initial research prototype to explore the abstractions and design tradeoffs of the proposed storage model. It currently supports the creation of databases, branches, and simple views, as well as transactions to update the relations. It does not support columnar storage; however, it does support partitioning the attributes of a relation into different groups (each must contain the primary key) that are stored in different sets of chunks.
 The prototype currently implements content-addressable storage using SHA-256 hashes and supports both CAS-based structures and B+-tree-style chunking. Several planned features are in the process of being implemented, including columnar layout within chunks and lazy materialization of chunk contents. In addition, many storage parameters are currently configured manually rather than being automatically tuned.

We conducted a small set of preliminary experiments to evaluate sharing behavior across different chunking strategies and workloads. We use a single table with 5 attributes and 50000 rows initially bulk-loaded. We then execute
sequences of 500 commits, each consisting of 200 inserts/updates/deletes. The experiments compare two chunking strategies: a capacity-based policy resembling B+-tree splitting
with fixed chunk sizes, and a content-based policy inspired by Prolly trees using a rolling hash over primary key values to determine chunk boundaries. Both strategies are configured to produce similar average chunk sizes to enable a fair comparison.

We evaluate several workload patterns, including append-only inserts, localized updates that concentrate modifications around nearby keys, uniformly distributed updates, and a mixed workload containing inserts, updates, and deletes. To study the effect of attribute grouping on sharing behavior, we also compare a row-oriented layout in which all attributes are stored together with a configuration that separates columns into two attribute groups maintained in independent chunk trees.

Table~\ref{tab:chunking} summarizes representative results illustrating how chunking strategy influences structural sharing across versions and
branches. In particular, content-based chunking tends to preserve chunk boundaries under localized modifications, resulting in greater reuse of chunks across versions compared to fixed-capacity splitting. However, the average
chunk size is significantly higher for content-based chunking which results in larger overall storage requirements when the updates are scattered (e.g., with {\em uniform updates} workload). In another set of experiments, we
created 5 branches from the same initial version, and did 250 commits on each with a mixed workload. Content-based chunking resulted in 1.5M chunks vs 3M chunks for capacity-based chunking, with the former requiring 50\% less
storage.

\begin{table}[t]
\centering
\caption{Effect of chunking strategy on structural sharing across workloads. For each setting, we report: (the number of unique chunks, total storage).}
\label{tab:chunking}
\begin{tabular}{lcc}
\toprule
Workload & Capacity-based Chunking & Content-based Chunking \\
\midrule
Append-only &  (1.2M, 1.1GB)& (19k, 35MB) \\
Localized update & (88k, 90MB) & (41.5k, 155MB) \\
Uniform update &(129k, 161MB)  & (90k, 409MB) \\
Mixed workload &(1.4M, 1.44GB)  & (650k, 1GB) \\
\bottomrule
\end{tabular}
\end{table}

We also evaluated the impact of column-store-style attribute grouping on structural sharing by comparing two configurations: a row-store layout where all columns 
are stored in a single chunk tree, and a column-store layout with two independent trees, each containing the primary key plus a disjoint subset of value columns. 
We simulated a workload of 500 commits against a 50000-row table, where odd-numbered commits update only first two columns, and even-numbered commits update the latter two 
(200 uniformly random updates per commit). Under capacity-based chunking, the column-store layout reduced total storage from 160.6 MB to 130.4 MB (18.8\% reduction), 
because the attribute group containing untouched columns rebuilds with identical data and produces the same content hashes, achieving full deduplication in the chunk store. 
The effect is more pronounced under content-based chunking, where column-store grouping reduces storage from 410.3 MB to 263.8 MB (35.7\% reduction). 
These results demonstrate that decomposing wide tables into attribute groups aligned with update patterns can substantially reduce the storage overhead of maintaining versioned 
data, and that this technique is orthogonal to and composable with the choice of splitting strategy.

\section{Conclusion and Future Work}
Our work started out as an investigation into how to best support schema evolution and versioning as first-class constructs in a modern database, to address many pain points we
see practitioners dealing with in practice. However, we realized that the abstraction that combines the two naturally encompasses other types of data transformations,
including views, continuous queries, and derived objects (like models). Currently each of these is typically managed in a separate manner--schema evolution is often handled in
an ad hoc manner or in a layer above the database (e.g., ORMs); limited types of view maintenance is supported inside the database but external cron jobs may need to be used for some use cases; temporal versioning is increasingly built-in but branching is handled in an ad hoc manner; and so on. Combining these into a single abstraction, even if the
abstraction is somewhat complex, will not only make it easier to use these constructs but also naturally enables additional functionality (e.g., versioning of views, temporal
or otherwise, is not supported in any system to our knowledge). 

There are many research challenges in realizing the vision of such a system. We are currently investigating these issues in the context of ErbiumDB, an entity-relationship
database that we are building~\cite{erbium}. In particular, we are investigating the best storage structures to support the key functions of the abstraction, as well as
concurrency control issues in maintaining multiple live branches that are bidirectionally or uni-directionally synced. The abstraction itself, however, is more general and there
are many other novel challenges in supporting it in other data management systems and data lakes.

\section*{AI-Generated Content Acknowledgement}
The author used AI-based tools, including OpenAI ChatGPT and Google Gemini, to assist with refining and editing portions of the manuscript text for clarity and presentation. These tools were used only for language refinement and did not generate the core technical ideas, results, or conclusions of the work. In addition, AI-assisted coding tools, specifically Claude Code, were used during the development of the research prototype to support code generation and implementation tasks.

\bibliographystyle{alpha}
\bibliography{main}

\end{document}